\documentclass[
    10pt,
    twocolumn,
    secnumarabic,
    amssymb, 
    nobibnotes, 
    aps, 
    prl,
    longbibliography,
    superscriptaddress
    ]{revtex4-2}

\usepackage[utf8]{inputenc}
\usepackage{siunitx}
\DeclareSIUnit{\echarge}{\ensuremath{\mathit{e}}}
\usepackage{tikz}
\usepackage{natbib}
\usepackage{graphicx}
\usepackage{appendix}
\usepackage{amsmath} 
\usepackage{booktabs}      
\usepackage{multirow}      
\usepackage{nicefrac}      
\usepackage{hyperref}

\date{\today}

\newcommand{\ie}{\textit{i.e.} }

\newcommand{\phirel}{\varphi_\mathrm{rel} }
\newcommand{\phirelzero}{\varphi_{\mathrm{rel,0}}}
\newcommand{\fc}{f_\mathrm{rev} }

\newcommand{\pv}{p_\mathrm{v} }
\newcommand{\ph}{p_\mathrm{h} }

\newcommand{\xiedm}{\xi_{\text{EDM}}} 

\begin{document}

\title{First Experimental Limit on the Permanent Electric Dipole Moment of the Deuteron}


\author{A.\,Andres}
\thanks{Present address: GSI Helmholtzzentrum für Schwerionenforschung GmbH,
Planckstr. 1, 64291 Darmstadt, Germany}
\affiliation{III. Physikalisches Institut B, RWTH Aachen University, 52056 Aachen, Germany}
\affiliation{Institut f\"ur Kernphysik, Forschungszentrum J\"ulich, 52425 J\"ulich, Germany}

\author{V.\,Hejny}
\affiliation{Institut f\"ur Kernphysik, Forschungszentrum J\"ulich, 52425 J\"ulich, Germany}

\author{A.\,Nass}
\affiliation{Institut f\"ur Kernphysik, Forschungszentrum J\"ulich, 52425 J\"ulich, Germany}

\author{N.N.\,Nikolaev}
\thanks{Present address: N.N. Bogoliubov Laboratory of Theoretical Physics, 141980 Dubna, Russia}
\affiliation{L.D. Landau Institute for Theoretical Physics, 142432 Chernogolovka, Russia}

\author{J.\,Pretz}
\affiliation{III. Physikalisches Institut B, RWTH Aachen University, 52056 Aachen, Germany}
\affiliation{Institut f\"ur Kernphysik, Forschungszentrum J\"ulich, 52425 J\"ulich, Germany}	

\author{F.\,Rathmann}
\thanks{Present address: Brookhaven National Laboratory, Upton, NY 11973, USA}
\affiliation{Institut f\"ur Kernphysik, Forschungszentrum J\"ulich, 52425 J\"ulich, Germany}

\author{V.\,Shmakova}
\thanks{Present address: Brookhaven National Laboratory, Upton, NY 11973, USA}
\affiliation{University of Ferrara and Istituto Nazionale di Fisica Nucleare, 44100 Ferrara, Italy}

\author{J. Slim}
\thanks{Present address: Deutsches Elektronen-Synchrotron, 22607 Hamburg, Germany}
\affiliation{III. Physikalisches Institut B, RWTH Aachen University, 52056 Aachen, Germany}

\author{F.\,Abusaif}
\thanks{Present  address: Karlsruhe Institute of Technology, Hermann-von-Helmholtz-Platz 1, 76344 Eggenstein-Leopoldshafen,
	Germany}
\affiliation{III. Physikalisches Institut B, RWTH Aachen University, 52056 Aachen, Germany}
\affiliation{Institut f\"ur Kernphysik, Forschungszentrum J\"ulich, 52425 J\"ulich, Germany}

\author{A.\,Aggarwal}
\affiliation{Marian Smoluchowski Institute of Physics, Jagiellonian University, 30348 Cracow, Poland}

\author{A.\,Aksentev}
\affiliation{Institute for Nuclear Research, Russian Academy of Sciences, 117312 Moscow, Russia}

\author{B.\,Alberdi}
\thanks{Present address: Humboldt-Universität zu Berlin, Institut für Physik, Newton-Straße 15, 12489 Berlin, Germany}
\affiliation{III. Physikalisches Institut B, RWTH Aachen University, 52056 Aachen, Germany}
\affiliation{Institut f\"ur Kernphysik, Forschungszentrum J\"ulich, 52425 J\"ulich, Germany}	

\author{L.\,Barion}
\affiliation{University of Ferrara and Istituto Nazionale di Fisica Nucleare, 44100 Ferrara, Italy}

\author{I.\,Bekman}
\thanks{Present address: Institute of Technology and Engineering, Forschungszentrum J\"ulich, 52425 J\"ulich, Germany}
\affiliation{Institut f\"ur Kernphysik, Forschungszentrum J\"ulich, 52425 J\"ulich, Germany}

\author{M.\, Bey\ss}
\affiliation{III. Physikalisches Institut B, RWTH Aachen University, 52056 Aachen, Germany}
\affiliation{Institut f\"ur Kernphysik, Forschungszentrum J\"ulich, 52425 J\"ulich, Germany}

\author{C.\,B\"ohme}
\affiliation{Institut f\"ur Kernphysik, Forschungszentrum J\"ulich, 52425 J\"ulich, Germany}

\author{B.\,Breitkreutz}
\affiliation{Institut f\"ur Kernphysik, Forschungszentrum J\"ulich, 52425 J\"ulich, Germany}

\author{N.\,Canale}
\affiliation{University of Ferrara and Istituto Nazionale di Fisica Nucleare, 44100 Ferrara, Italy}

\author{G.\,Ciullo}
\affiliation{University of Ferrara and Istituto Nazionale di Fisica Nucleare, 44100 Ferrara, Italy}

\author{S.\,Dymov}
\affiliation{University of Ferrara and Istituto Nazionale di Fisica Nucleare, 44100 Ferrara, Italy}

\author{N.-O.\, Fr\"ohlich}
\thanks{Present address: Deutsches Elektronen-Synchrotron, 22607 Hamburg, Germany}
\affiliation{Institut f\"ur Kernphysik, Forschungszentrum J\"ulich, 52425 J\"ulich, Germany}

\author{R.\,Gebel}
\affiliation{Institut f\"ur Kernphysik, Forschungszentrum J\"ulich, 52425 J\"ulich, Germany}

\author{M.\,Gaisser}
\affiliation{III. Physikalisches Institut B, RWTH Aachen University, 52056 Aachen, Germany}

\author{K.\,Grigoryev}
\thanks{Present address: GSI Helmholtzzentrum für Schwerionenforschung GmbH,
Planckstr. 1, 64291 Darmstadt, Germany}
\affiliation{Institut f\"ur Kernphysik, Forschungszentrum J\"ulich, 52425 J\"ulich, Germany}

\author{D.\,Grzonka}
\affiliation{Institut f\"ur Kernphysik, Forschungszentrum J\"ulich, 52425 J\"ulich, Germany}

\author{D.\,Gu}
\affiliation{III. Physikalisches Institut B, RWTH Aachen University, 52056 Aachen, Germany}
\affiliation{GSI Helmholtzzentrum für Schwerionenforschung GmbH,
Planckstr. 1, 64291 Darmstadt, Germany}

\author{D.\,Heberling}
\affiliation{Institut für Hochfrequenztechnik, RWTH Aachen University, 52056 Aachen, Germany}

\author{J.\, Hetzel}
\thanks{Present address: GSI Helmholtzzentrum für Schwerionenforschung GmbH,
Planckstr. 1, 64291 Darmstadt, Germany}
\affiliation{Institut f\"ur Kernphysik, Forschungszentrum J\"ulich, 52425 J\"ulich, Germany}

\author{D. H\"olscher}
\affiliation{Institut für Hochfrequenztechnik, RWTH Aachen University, 52056 Aachen, Germany}

\author{O.\,Javakhishvili}
\thanks{Present address: Faculty of Nuclear Sciences and Physical Engineering, Czech Technical University in Prague, 160 00 Praha 6, Czech Republic}
\affiliation{Department of Electrical and Computer Engineering, Agricultural University of Georgia, 0159 Tbilisi, Georgia}

\author{A.\,Kacharava}
\affiliation{Institut f\"ur Kernphysik, Forschungszentrum J\"ulich, 52425 J\"ulich, Germany}

\author{V.\,Kamerdzhiev}
\thanks{Present address: GSI Helmholtzzentrum für Schwerionenforschung GmbH,
Planckstr. 1, 64291 Darmstadt, Germany}
\affiliation{Institut f\"ur Kernphysik, Forschungszentrum J\"ulich, 52425 J\"ulich, Germany}

\author{S.\,Karanth}
\thanks{Present address: H. Niewodnicza\'{n}ski Institute of Nuclear Physics, Polish Academy of Sciences, 31342, Krak\'{o}w, Poland}
\affiliation{Marian Smoluchowski Institute of Physics, Jagiellonian University, 30348 Cracow, Poland}

\author{I.\,Keshelashvili}
\thanks{Present address: GSI Helmholtzzentrum für Schwerionenforschung GmbH,
Planckstr. 1, 64291 Darmstadt, Germany}
\affiliation{Institut f\"ur Kernphysik, Forschungszentrum J\"ulich, 52425 J\"ulich, Germany}

\author{A.\,Kononov}
\affiliation{University of Ferrara and Istituto Nazionale di Fisica Nucleare, 44100 Ferrara, Italy}

\author{K.\,Laihem}
\thanks{Present address: GSI Helmholtzzentrum für Schwerionenforschung GmbH,
Planckstr. 1, 64291 Darmstadt, Germany}
\affiliation{III. Physikalisches Institut B, RWTH Aachen University, 52056 Aachen, Germany}

\author{A.\,Lehrach}
\affiliation{III. Physikalisches Institut B, RWTH Aachen University, 52056 Aachen, Germany}
\affiliation{Institut f\"ur Kernphysik, Forschungszentrum J\"ulich, 52425 J\"ulich, Germany}

\author{P.\,Lenisa}
\affiliation{University of Ferrara and Istituto Nazionale di Fisica Nucleare, 44100 Ferrara, Italy}

\author{N.\,Lomidze}
\affiliation{High Energy Physics Institute, Tbilisi State University, 0186 Tbilisi, Georgia}

\author{B.\,Lorentz}
\affiliation{GSI Helmholtzzentrum für Schwerionenforschung GmbH,
Planckstr. 1, 64291 Darmstadt, Germany}

\author{G.\,Macharashvili}
\affiliation{High Energy Physics Institute, Tbilisi State University, 0186 Tbilisi, Georgia}

\author{A.\,Magiera}
\affiliation{Marian Smoluchowski Institute of Physics, Jagiellonian University, 30348 Cracow, Poland}

\author{M.\,Margos}
\affiliation{III. Physikalisches Institut B, RWTH Aachen University, 52056 Aachen, Germany}
\affiliation{GSI Helmholtzzentrum für Schwerionenforschung GmbH,
Planckstr. 1, 64291 Darmstadt, Germany}

\author{D.\,Mchedlishvili}
\affiliation{High Energy Physics Institute, Tbilisi State University, 0186 Tbilisi, Georgia}

\author{A.\,Melnikov}
\affiliation{Institute for Nuclear Research, Russian Academy of Sciences, 117312 Moscow, Russia}

\author{F.\,Müller}
\affiliation{III. Physikalisches Institut B, RWTH Aachen University, 52056 Aachen, Germany}
\affiliation{Institut f\"ur Kernphysik, Forschungszentrum J\"ulich, 52425 J\"ulich, Germany}

\author{D.\,Okropiridze}
\thanks{Present address: Ruhr-Universit\"at Bochum, Institut f\"ur Experimentalphysik I, 44801 Bochum, Germany}
\affiliation{High Energy Physics Institute, Tbilisi State University, 0186 Tbilisi, Georgia}

\author{A.\,Pesce}
\affiliation{Institut f\"ur Kernphysik, Forschungszentrum J\"ulich, 52425 J\"ulich, Germany}

\author{A.\,Piccoli}
\affiliation{University of Ferrara and Istituto Nazionale di Fisica Nucleare, 44100 Ferrara, Italy}

\author{V.\,Poncza}
\affiliation{Institut f\"ur Kernphysik, Forschungszentrum J\"ulich, 52425 J\"ulich, Germany}

\author{D.\,Prasuhn}
\affiliation{Institut f\"ur Kernphysik, Forschungszentrum J\"ulich, 52425 J\"ulich, Germany}

\author{A.\,Saleev}
\thanks{Present address: Institut f\"ur nachhaltige Wasserstoffwirtschaft, Forschungszentrum J\"ulich, 52425 J\"ulich, Germany}
\affiliation{University of Ferrara and Istituto Nazionale di Fisica Nucleare, 44100 Ferrara, Italy}

\author{D.\,Shergelashvili}
\affiliation{High Energy Physics Institute, Tbilisi State University, 0186 Tbilisi, Georgia}

\author{R.\,Shankar}
\affiliation{University of Ferrara and Istituto Nazionale di Fisica Nucleare, 44100 Ferrara, Italy}

\author{N.\,Shurkhno}
\thanks{Present address: GSI Helmholtzzentrum für Schwerionenforschung GmbH,
Planckstr. 1, 64291 Darmstadt, Germany}
\affiliation{Institut f\"ur Kernphysik, Forschungszentrum J\"ulich, 52425 J\"ulich, Germany}

\author{S.\,Siddique}
\thanks{Present address: GSI Helmholtzzentrum für Schwerionenforschung GmbH,
Planckstr. 1, 64291 Darmstadt, Germany}
\affiliation{III. Physikalisches Institut B, RWTH Aachen University, 52056 Aachen, Germany}
\affiliation{Institut f\"ur Kernphysik, Forschungszentrum J\"ulich, 52425 J\"ulich, Germany}

\author{A.\,Silenko}
\affiliation{Bogoliubov Laboratory of Theoretical Physics, International Intergovernmental Organization Joint Institute for Nuclear Research, 141980 Dubna, Russia}

\author{H.\,Soltner}
\affiliation{Institute of Technology and Engineering, Forschungszentrum J\"ulich, 52425 J\"ulich, Germany}

\author{R.\,Stassen}
\affiliation{Institut f\"ur Kernphysik, Forschungszentrum J\"ulich, 52425 J\"ulich, Germany}

\author{E.J.\,Stephenson}		
\affiliation{Indiana University, Department of Physics, Bloomington, Indiana 47405, USA}

\author{H.\,Ströher}
\affiliation{Institut f\"ur Kernphysik, Forschungszentrum J\"ulich, 52425 J\"ulich, Germany}

\author{M.\,Tabidze}
\affiliation{High Energy Physics Institute, Tbilisi State University, 0186 Tbilisi, Georgia}

\author{G.\,Tagliente}
\affiliation{Istituto Nazionale di Fisica Nucleare sez.\ Bari, 70125 Bari, Italy}

\author{V.\,Tempel}
\affiliation{III. Physikalisches Institut B, RWTH Aachen University, 52056 Aachen, Germany}
\affiliation{GSI Helmholtzzentrum für Schwerionenforschung GmbH,
Planckstr. 1, 64291 Darmstadt, Germany}

\author{Y.\,Valdau}
\thanks{Present address: GSI Helmholtzzentrum für Schwerionenforschung GmbH,
Planckstr. 1, 64291 Darmstadt, Germany}
\affiliation{Institut f\"ur Kernphysik, Forschungszentrum J\"ulich, 52425 J\"ulich, Germany}

\author{M.\,Vitz}
\affiliation{III. Physikalisches Institut B, RWTH Aachen University, 52056 Aachen, Germany}
\affiliation{Institut f\"ur Kernphysik, Forschungszentrum J\"ulich, 52425 J\"ulich, Germany}

\author{T.\,Wagner}
\thanks{Present address: GSI Helmholtzzentrum für Schwerionenforschung GmbH,
Planckstr. 1, 64291 Darmstadt, Germany}
\affiliation{III. Physikalisches Institut B, RWTH Aachen University, 52056 Aachen, Germany}
\affiliation{Institut f\"ur Kernphysik, Forschungszentrum J\"ulich, 52425 J\"ulich, Germany}

\author{A.\,Wirzba}
\affiliation{Institute for Advanced Simulation, Forschungszentrum J\"ulich, 52425 J\"ulich, Germany}

\author{A.\,Wro\'{n}ska}
\affiliation{Marian Smoluchowski Institute of Physics, Jagiellonian University, 30348 Cracow, Poland}

\author{P.\,W\"ustner}
\affiliation{Institute of Technology and Engineering, Forschungszentrum J\"ulich, 52425 J\"ulich, Germany}

\author{M. \.{Z}urek}
\thanks{Present address: Argonne National Laboratory, Lemont, Illinois 60439, USA}
\affiliation{Institut f\"ur Kernphysik, Forschungszentrum J\"ulich, 52425 J\"ulich, Germany}

\collaboration{JEDI Collaboration}
\noaffiliation

\begin{abstract}
    Permanent electric dipole moments (EDMs) provide a sensitive probe of physics beyond the Standard Model and are directly linked to additional sources of CP violation that could explain the matter-antimatter asymmetry of the universe. EDM measurements of charged particles in storage rings rely on detecting a small tilt of the invariant spin axis with respect to the ring plane. In this work, we present the experimental determination of the invariant spin axis of an ensemble of deuterons in the COoler SYnchrotron (COSY), a conventional magnetic storage ring, using a combination of a radio-frequency Wien filter, a superconducting Siberian snake and an electron-cooler solenoid. The measurements reveal tilts of a few milliradians, which are dominated by systematic effects. From the observed tilts, we derive the first experimental limit on the deuteron EDM, $|d^d|<\SI{2.5e-17}{\echarge\cdot cm} \; (\SI{95}{\%} \text{ C.L.})$. This result demonstrates the feasibility of using storage rings to search for EDMs of charged stable hadrons and provides a foundation for future dedicated facilities.
\end{abstract}

\maketitle

Permanent electric dipole moments (EDMs) of elementary and composed sub-atomic particles violate both parity (P) and time-reversal (T) symmetry~\cite{Khriplovich:1997ga,Pospelov:2005pr}. 
If the combined CPT symmetry -- where C represents charge conjugation -- is conserved, then T violation implies CP violation~\cite{Luders:1957zz,Greenberg:2002uu}. 
So far CP violation has been observed, for example, in the decays of neutral kaons~\cite{PhysRevLett.13.138}, B mesons~\cite{Gershon_2017,Aubert:2001nu} and baryons~\cite{LHCb:2025ray}. 
The Standard Model predicts EDMs so tiny that their detection is beyond the current experimental reach~\cite{Engel:2013lsa,chupp:2017rkp}. 
Consequently, any measurable EDM would be a hint at the presence of physics beyond the Standard Model (or a non-zero QCD $\theta$ term)~\cite{Pospelov:2005pr}. 
Therefore, EDMs are considered to be sensitive probes to search for additional sources of CP violation, which are needed to explain the observed matter-antimatter asymmetry in the universe~\cite{Sakharov:1967dj}. 
This paper discusses the first measurement of the deuteron EDM building on a sequence of milestones previously achieved and published by the JEDI collaboration~\cite{Guidoboni:2016bdn,Eversmann:2015jnk,PhysRevAccelBeams.20.072801,PhysRevResearch.7.023257,PhysRevAccelBeams.28.062801}. 

\begin{figure}
    \centering
    \includegraphics[width=\linewidth]{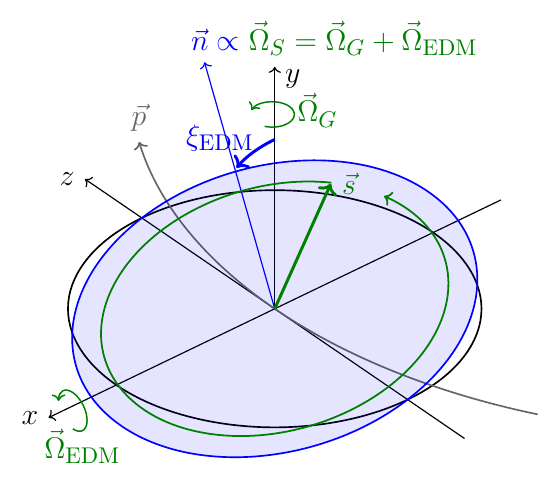}
    \caption{Influence of the EDM on the invariant spin axis $\vec n$. The gray line marks the beam trajectory. Assuming an idealized storage ring and no EDM, the spin is precessing in the $x$-$z$ plane. A non-zero EDM tilts the spin-precession plane around the $z$-axis by an angle $\xiedm$ indicated by the blue plane. The situation shown here corresponds to a particle with $G<0$, e.g. a deuteron.
    The spin $\vec s$ precesses in opposite direction to the momentum $\vec p$. For a positive EDM, i.e. $\eta >0$, the angle $\xiedm$ defined in Eq.~(\ref{eq:eta_edm}) is negative.} \label{fig:ISA_rotation}
\end{figure}

In general, an EDM of a sub-atomic particle can be measured by studying its influence on the particle's spin motion, since the EDM is aligned with its spin direction. 
Quantitatively, the spin dynamics is governed by the Thomas-BMT equation~\cite{Bargmann:1959gz,Nelson:1959zz,Fukuyama:2013ioa}.
In a purely magnetic storage ring with a vertical magnetic field and $\vec\beta \perp \vec B$, the spin evolution relative to the momentum vector can be described in a co-rotating frame.
Subtracting the cyclotron rotation removes all kinematic and Dirac ($g=2$) contributions, such that the remaining spin precession is governed solely by the anomalous magnetic moment $G$ and a possible EDM (see, \emph{e.g.},\cite{chupp:2017rkp}).
Neglecting terms proportional to $\vec\beta\cdot\vec B$, the resulting equation of motion reads:
\begin{eqnarray}
\frac{\text{d} \vec{S}}{\text{d}t} = \vec\Omega_S \times \vec{S} &=& (\vec{\Omega}_G + \vec{\Omega}_{\mathrm{EDM}}) \times \vec{S},  \label{eq:tbmt}\\  
 \vec{\Omega}_G &=&  -\frac{q}{m} ~ G \vec B \label{eq:tbmt_mdm},\\
 \vec{\Omega}_{\mathrm{EDM}} &=& -\frac{q}{m} ~\frac{\eta}{2} \,  \vec{\beta} \times \vec{B}. \label{eq:omedm}
\end{eqnarray}
Here, $\vec\Omega_G$ denotes exclusively the anomalous magnetic-moment-induced spin precession
and $\vec{\Omega}_{\mathrm{EDM}}$ the EDM-induced spin rotation.
$\vec{S}$ is the spin vector in units of $\hbar$ in the particle rest frame, $t$ the time in the laboratory system, $m$ and $q$ the rest mass and the charge of the particle, $\beta$ the particle velocity in units of $c$ and $\vec{B}$ the magnetic field in the laboratory system. 
The dimensionless quantities $G$ and $\eta$ are related to the magnetic dipole moment (MDM) $\vec \mu$ and the EDM $\vec d$:
\begin{equation}
\vec{\mu} = 
g \frac{q \hbar}{2 m} \vec{S} = 
(1+G) \frac{q \hbar}{m} \vec{S}\, \quad \mbox{and} \quad \vec{d} = \eta \frac{q \hbar}{2 m c} \vec{S} \, .
\label{eq:mdm_and_edm}
\end{equation}
With the given assumptions, the anomalous magnetic moment and an EDM cause precessions around the vertical $y$-axis and the radial $x$-axis, respectively (see Fig.~\ref{fig:ISA_rotation}):
\begin{alignat}{2}
\vec{\Omega}_G &{}={}& \Omega_G&\;\vec{e}_\mathrm{y},  \\
\vec{\Omega}_\mathrm{EDM} &{}={}& \Omega_\mathrm{EDM}&\;\vec{e}_\mathrm{x}.
\end{alignat}
It is convenient to introduce the invariant spin axis $\vec n$ as a unit vector normal to the tilted precession plane, \ie $\vec\Omega_S = \Omega_S\,\vec{n}$, and its tilt angle $\xiedm$ in the $x$-$y$ plane such that 
\begin{equation}\label{eq:n}
\vec n=\left(-\sin\xiedm,\cos\xiedm,0 \right).
\end{equation}
With $|\eta|\ll 1$ one gets to first order in $\eta$
\begin{equation}
\label{eq:eta_edm}
\xiedm=-\frac{\Omega_{\mathrm{EDM}}}{\Omega_G} = \frac{\eta \beta}{2 G}.
\end{equation}
Solving Eq. (\ref{eq:tbmt}) one finds for a spin vector initially pointing in longitudinal direction ($\vec S(0) = (0,0,1)$) to first order in $\eta$ 
\begin{equation}\label{eq:St}
\vec S(t) = \left(\sin(\Omega_G t), \xiedm \sin(\Omega_G t) ,\cos(\Omega_G t)\right),
\end{equation}
\ie the EDM causes an oscillation of the vertical polarization component with amplitude $|\xiedm|$. This signature is used in the muon $g-2$ experiment to measure the muon EDM \cite{PhysRevD.80.052008}.
For deuterons this approach is far less sensitive because the magnitude of the magnetic anomaly $G$ is much larger than that of the muon.

To induce a macroscopic build-up of a vertical polarization an additional radio-frequency (rf) Wien filter \cite{Rathmann:2013rqa,PhysRevSTAB.16.114001} in resonance with the spin precession is applied such that the Lorentz force vanishes \cite{Slim:2016pim} and the electric and magnetic fields only act on the spin vector. 
For a bunched beam, the rf Wien filter applies a spin kick turn by turn and is operated at a spin-resonance frequency corresponding to the $k^\textrm{th}$ harmonic of the revolution frequency.
The additional spin rotation around the magnetic field axis $\vec{m}$ of the rf Wien filter per turn is
\begin{eqnarray}
    \psi(t) &=& \psi_0 \sin(\Omega_\mathrm{WF} t + \phirel) \quad \mbox{with} \label{eq:wfrot}\\
    \psi_0 &=& \frac{q}{m} \frac{G+1}{\gamma^2\beta c} \int B_\textrm{WF}\,\text{d}l, \\
    \Omega_\mathrm{WF} &=& | k\,\Omega_\mathrm{rev} + \Omega_S | \,,\quad k\in\mathbb{Z}\,,
\end{eqnarray}
where $B_\textrm{WF}$ and $\Omega_\mathrm{WF}$ denote the magnetic field  and the radio frequency of the rf Wien filter, respectively, and $\phirel$ is the relative phase between the radio frequency and the precession in the horizontal plane. 
The corresponding spin dynamics is presented in detail in Refs.~\cite{Rathmann:2019lwi,PhysRevAccelBeams.27.111002}. Below, only the key elements are summarized.

Solving Eq.~(\ref{eq:tbmt}) for this case, the fast vertical oscillation in Eq.~(\ref{eq:St}) is superimposed by a slow oscillation (also see Fig.~10 of Ref.~\cite{PhysRevAccelBeams.28.062801})
\begin{equation}\label{eq:oscwf}
S_y(t) = \cos(\phirel)\sin \left( 2\pi f_y t \right) .
\end{equation}
The frequency $f_y$ depends on the relative orientation of the magnetic-field axis $\vec m$ and the static invariant spin axis $\vec n$ in the absence of the rf excitation.
Aligned axes do not induce an oscillation, whereas a perpendicular configuration yields the maximal effect. If $\vec{m}$ is vertical, this sensitivity allows the rf Wien filter to probe the tilt of the invariant spin axis. 
As experimental observable one can then define the resonance tune (or resonance strength) directly related to $\xiedm$:
\begin{equation}
    \epsilon = \frac{f_y}{\fc} = \frac{\psi_0}{4\pi}|\vec{n}\times\vec{m}| = \frac{\psi_0}{4\pi}|\sin(\xiedm)|.  \label{eq:resonance_strength}
\end{equation}

\begin{figure}
    \begin{center}
        \includegraphics[width=\linewidth]{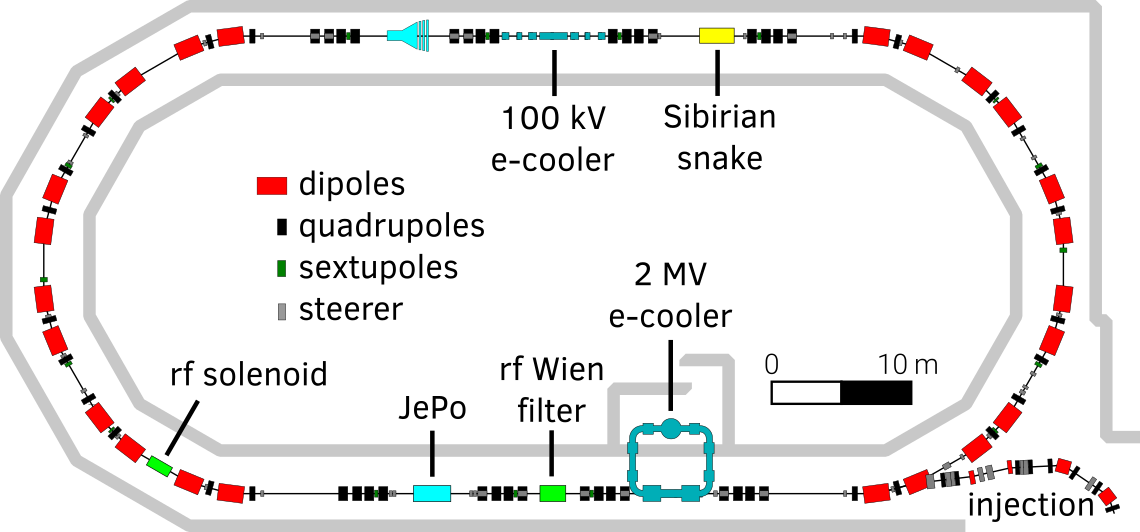}
        \caption{Sketch of the COSY ring indicating the position of the relevant installations: electron coolers, rf Wien filter, Siberian snake and the JEDI polarimeter (JePo).  
        \label{fig:cosy}}
    \end{center}
\end{figure}

This describes the ideal case where only the EDM contributes to the tilt of $\vec{n}$. In an actual ring, however, magnet misalignments, field imperfections, and orbit deviations introduce additional, position-dependent contributions $n_x^\mathrm{sys}$ and $n_z^\mathrm{sys}$
\begin{equation}
\vec n=\left(n_x^\mathrm{sys}-\xiedm,1,n_z^\mathrm{sys}\right) \;\mathrm{for}\; |\xiedm|,|n_x^\mathrm{sys}|,|n_z^\mathrm{sys}|\ll 1.
\end{equation}
Therefore, instead of aligning the rf Wien filter’s magnetic field axis $\vec{m}$ vertically and measuring only the build-up, we followed a different strategy. 
The rf Wien filter could be rotated around the beam axis ($\phi_{\text{WF}}$) to introduce an additional $x$-component (roll) to $\vec{m}$, while a spin rotation in a Siberian snake located on the opposite side of the ring (see Fig.~\ref{fig:cosy}) added a $z$-component (pitch) to $\vec{n}$ ($\phi_\mathrm{snake}$). 
For small additional tilts one gets 
\begin{eqnarray}
\vec{n} &=& (n_x, 1, n_z + \phi_\mathrm{snake}), \\ 
\vec{m} &=& (\phi_\mathrm{WF}, 1, 0)
\end{eqnarray}
and using Eq.~(\ref{eq:resonance_strength}) $\epsilon^2$ can be written in leading order as
\begin{equation}
  \epsilon^2 = \frac{\psi_0^2}{16\pi^2}\left[\left(n_x-\phi_{\text{WF}}\right)^{2}+\left(n_z + \phi_\mathrm{snake}\right)^{2}\right]. \label{eq:res_strength_sib}
\end{equation}

Eq.~(\ref{eq:res_strength_sib}) describes a two-dimensional paraboloid with the minimum corresponding to 
$\vec{m}\parallel\vec{n}$, where the resonance strength vanishes. The corresponding angles $\phi_{\text{WF}}^{\text{min}}$ and $\phi_\mathrm{snake}^{\text{min}}$ are a measure of the direction of the invariant spin axis at the rf Wien filter location such that $\vec{n} = (\phi_{\text{WF}}^{\text{min}},1,-\phi_\mathrm{snake}^{\text{min}})$.
This approach facilitates a systematic study of how experimental uncertainties affect the extracted parameters and the final result. 

\begin{table}[t]
\centering
\caption{Beam and machine parameters. The deuteron mass $m$ and the deuteron $g$ factor are taken from the NIST database~\cite{NIST_database}.}
\begin{tabular}{llrr}
\hline
\textbf{Parameter} & \textbf{Symbol} & \textbf{Unit} & \textbf{Value} \\
\hline
Deuteron momentum (lab) & $p$ & [MeV/c] & \SI{970}{}  \\
Particles per bunch &N& & $(0.5-1)\times 10^{9}$ \\
Lorentz factor & $\gamma$ & & 1.126  \\
Beam velocity & $\beta$ & & 0.46  \\
COSY circumference & $L_{\mathrm{COSY}}$ & [m] & \SI{183.6}{}  \\
Revolution frequency & $f_{\mathrm{rev}}$ & [Hz] & \SI{750602.6}{}  \\
Spin precession frequency & $f_s$ & [Hz] & \SI{120847.3}{}  \\
Deuteron mass & $m$ & [MeV/c$^2$] & \SI{1875.61}{} \\
Deuteron $g$ factor & $g$ && \SI{1.714025}{}  \\
Deuteron $G = (g - 2)/2$ & $G$ && \SI{-0.142987}{}  \\
rf Wien filter field integral & $\int B_{\text{WF}}\,\text{d}l$ & [$\mu$Tm] & $7.9-12.0$ \\
\hline
\end{tabular}
\label{tab:beam_params}
\end{table}

The experiment was performed at the COoler SYnchrotron (COSY) at Forschungszentrum Jülich, utilizing the standard COSY infrastructure along with the rf Wien filter \cite{Slim:2016pim}, a superconducting Siberian snake \cite{Lehrach:556296,Lehrach:826216}, the solenoid of the 2-MV e-cooler, and the JEDI Polarimeter (JePo) \cite{Mueller_2020} (see Fig.~\ref{fig:cosy}). 
Results from beam-based alignment~\cite{Wagner_2021} and a dedicated geodetic survey of the ring magnets 
were implemented following the experience gained in earlier experiments.
A summary of standard parameters is given in Table \ref{tab:beam_params}. Each cycle involved the injection of vertically (vector-) polarized deuterons into the COSY ring, bunching and accelerating the beam to \qty{970}{\MeV\per c}, electron cooling to reduce emittance
and orbit corrections. After starting the extraction onto a carbon target for polarization measurements the deuteron spins were rotated into the horizontal plane using an rf solenoid. The spin precession was continuously monitored, and the rf frequency and phase of the Wien filter was adjusted accordingly to keep a constant phase between the Wien filter rf and the spin precession phase (phase locking)~\cite{Eversmann:2015jnk,PhysRevAccelBeams.28.062801}.

The experiment began with spin coherence time optimization ~\cite{Guidoboni:2016bdn, Guidoboni2018}, followed by systematic studies of the vertical polarization buildup. The data acquisition system was the same as described in \cite{PhysRevAccelBeams.28.062801}. The primary observables were the detector rates in the four quadrants up, right, down, left of JePo together with the radio frequencies of the accelerator cavity and the rf Wien filter. The rates were used to calculate the time-dependent up-down and left-right asymmetry as a measure of the in-plane and vertical polarization, respectively.

The measurements were organized in maps, defined as two-dimensional scans over $\phi_\mathrm{WF}$ and $\phi_\mathrm{snake}$, each at various relative phases $\varphi_\mathrm{rel}$, with all other parameters held fixed for one map. For this,  the two methods discussed conceptually in \cite{PhysRevAccelBeams.28.062801} were used: 
(i) one or more bunches, such that the phase locking is probing the same bunch as reference that the rf Wien filter acts on (single-bunch method),
and (ii) using two bunches with the rf Wien filter selectively acting on one bunch 
while the other bunch remained undisturbed (pilot-bunch method)~\cite{PhysRevResearch.7.023257}. 

\begin{figure}
    \begin{center}
      \includegraphics[width=\linewidth]{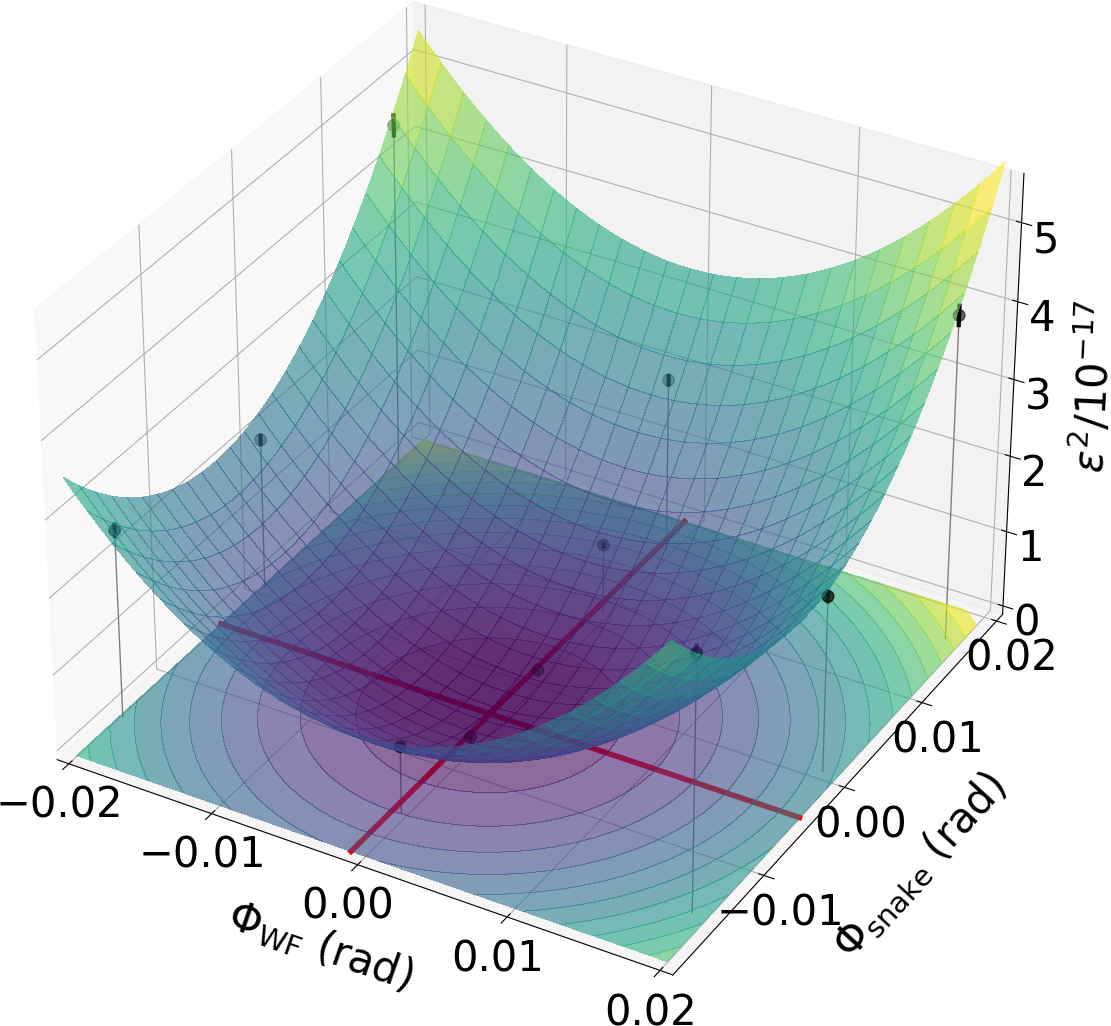}
        \caption{The resonance strength for Map 2/bunch 1 is shown as a function of the rf Wien filter rotation angle and the spin rotation in the Siberian snake. The minimum of the resulting two-dimensional paraboloid indicates the orientation of the invariant spin axis at the location of the rf Wien filter. 
        \label{fig:prec_1_map_2}}
    \end{center}
\end{figure}

In order to determine the resonance strength $\epsilon$, the in-plane and vertical polarizations were fitted simultaneously using model-based fit functions~\cite{Hempelmann:718035,PhysRevAccelBeams.27.111002}, with common parameters describing the resonance strength, oscillation amplitude, spin decoherence, and the deviation from resonance. The corresponding fit formulas are given in End Matter.
The squared resonance strength ($\epsilon^2$) is then plotted as a function of $\phi_\mathrm{WF}$ and $\phi_\mathrm{snake}$, as exemplarily shown in Fig.~\ref{fig:prec_1_map_2}. 
The minimum, and thus the direction of the invariant spin axis, is extracted by a parabolic fit using Eq.~(\ref{eq:res_strength_sib}). 

In total, seven maps were taken: the first two using the single bunch method and the remaining five using the pilot-bunch method. 
The measurement of each map took approximately 24 hours.
For systematic studies, the last two maps were taken with the 2\,MV e-cooler solenoid inducing an additional tilt of an invariant spin axis at the location of the rf Wien filter of $\Delta n_x = \pm \SI{1}{mrad}$ and $\Delta n_z = \pm \SI{1.8}{mrad}$; the plus signs apply to map 6 and the minus signs to map 7.
The results are summarized in Table \ref{tab:all_results}, the quoted uncertainties are statistical only.
The elevated reduced chi-squared values and the observed scatter reflect varying systematic offsets throughout the data, which dominate the chi-squared contributions at larger distances from the minimum.
Averaging yields
\begin{equation}
  n_{x,\text{avg.}} = \SI{-2.1(12)}{mrad}, \; n_{z,\text{avg.}} = \SI{3.9(6)}{mrad},
\end{equation}
where the uncertainties correspond to the RMS of the scatter of the results. 
These results are clearly dominated by systematic effects. Ideally, $n_x$ would be a measure of the EDM effect, while $n_z$ would be consistent with zero.

The experimental method relies on the magnetic-field axis of the rf Wien filter as a reference for determining the orientation of the invariant spin axis. 
While no direct experimental data exist on the magnetic-field direction within the rf Wien filter, extensive simulation studies have been performed, which explore the range of possible field errors perpendicular to the main axis~\cite{Slim:2016pim,Slim:2016dct,Slim:748558}. 
Additional simulations were performed that explicitly account for the geometry of the installed rf Wien filter, following the approach described in Ref.~\cite{Slim:2016dct}. 
The resulting possible spread corresponds to systematic uncertainties in the order of a few milliradians in the field orientation. The dominant contribution was found to originate from
a rotation of the beam axis in the horizontal plane. Further details are given in End Matter.

Extensive beam and spin-tracking simulations to study the influence of known misalignments of the ring magnets and the beam-position monitors and the corresponding uncertainties
have been carried out.
No longitudinal or radial components of the invariant spin axis larger than \SI{1}{mrad} could be found~\cite{Vitz:998208}.

As a further systematic check, the longitudinal component of the invariant spin axis at the locations of two solenoids was determined using the spin tune mapping~\cite{PhysRevAccelBeams.20.072801} developed by the JEDI collaboration.
In these measurements the rf Wien filter was not involved.
The results are
\begin{equation}
  n_{z,1} = \SI{-0.0565(7)}{mrad}, \; n_{z,2} = \SI{-0.0706(6)}{mrad},
\end{equation}
for the Siberian snake and the 2\,MV e-cooler solenoid, respectively. 
The longitudinal tilt below \SI{0.1}{mrad}, measured at the 2\,MV e-cooler solenoid only \SI{8}{m} upstream of the rf Wien filter, supports the conclusion that the main systematic uncertainties originate from the direction of the magnetic field axis of the rf Wien filter.

\begin{table}
\centering
\caption{Summary of all maps of the invariant spin axis. Maps 6 and 7 were taken with the 2\,MV e-cooler solenoid inducing an additional, fixed tilt of the invariant spin axis at the location of the rf Wien filter. The corresponding shift of the minimum has been taken into account.}
\begin{tabular}{llllll}
\toprule
Map & $n_x$/mrad & $n_z$/mrad & $\chi^2$/ndf & Method \\
\midrule
1 & $-1.72(12)$ & $4.90(6)$  & $455.5/8$ = 56.9 & single bunch\\
2 & $-1.34(11)$ & $4.18(4)$  & $147.0/8$ = 18.4 & single bunch\\
3 & $-3.29(36)$ & $3.54(31)$ & $16.5/5$ =  3.3 & pilot bunch \\
4 & $-1.96(8)$  & $3.81(4)$  & $163.8/21$ = 7.8 & pilot bunch \\
5 & $-2.50(8)$  & $3.61(5)$  & $201.8/21$ = 9.6 & pilot bunch \\
6 & $-0.61(12)$ & $3.15(6)$  & $52.5/5$ = 10.5 & pilot bunch \\
7 & $-4.84(17)$ & $3.35(10)$ & $127.5/5$ = 25.5 & pilot bunch \\
\bottomrule
\end{tabular}
\label{tab:all_results}
\end{table}

In an ideal storage ring, the invariant spin axis has no longitudinal component. As the observed longitudinal tilt is therefore predominantly caused by systematic effects, the measured tilt provides a direct estimate of the experimental uncertainty. This uncertainty can then be transferred to the radial component in order to construct an upper limit for the corresponding tilt:
\begin{equation}
     \sigma_{n_z^{\text{sys.}}} \approx \sigma_{n_x^{\text{sys.}}} \;\hat{=}\; n_{z,\text{avg.}} = \SI{3.9}\;{\text{mrad}}.
\end{equation}
Assuming that our measurement is consistent with an EDM value of zero and taking $\sigma_{n_z^{\text{sys.}}}=\SI{3.9}{mrad}$ as a one standard deviation Gaussian error for $n_x$, a 95\% CL can be obtained by multiplying by the corresponding factor 1.96.
Using Eqs. (\ref{eq:n}) and (\ref{eq:eta_edm}), this translates into an upper limit of the dimensionless EDM strength
\begin{equation}
    |\eta|<0.00475 \; (\SI{95}{\%} \text{ C.L.}).
\end{equation}
Consequently, a first limit on the deuteron EDM can be determined using Eq. (\ref{eq:mdm_and_edm})
\begin{equation}
    |d^d|<\SI{2.5e-17}{\echarge\cdot cm} \; (\SI{95}{\%} \text{ C.L.}).
\end{equation}

To put this result into perspective, the only measurement of a permanent EDM performed in a storage ring so far was carried out for muons, yielding an upper limit of 
 $|d^{\mu}|<\SI{1.8e-19}{e\cdot cm}\;(\SI{95}{\%}\text{ C.L.})$~\cite{PhysRevD.80.052008}.
 The sensitivity is governed by the ratio of the EDM to the magnetic anomaly, $G\approx-0.14$ for deuterons and $a \simeq 0.001$ for muons, resulting in about two orders of magnitude higher sensitivity for the muon. 
Furthermore, this result was obtained using a dedicated muon storage ring specifically designed for precision experiments. It should also be noted that at COSY a significantly more stringent limit has recently been obtained for an axion-field-induced oscillating EDM of the deuteron~\cite{PhysRevX.13.031004}. Owing to the resonant coupling to the axion field, most systematic effects cancel, illustrating the potential of storage-ring techniques.

This measurement demonstrates  that a storage ring can be used to access the EDM of charged hadrons by precisely determining the invariant spin axis.
At the present stage, the observed tilt of the invariant spin axis is still dominated by systematic effects, such as magnet misalignments, deviations from an ideal orbit, and field errors. 
This calls for further experimental studies aimed at reducing and validating the associated systematic uncertainties. An improved rf Wien filter setup, incorporating additional ferrites to enhance field strength and homogeneity, had already been prepared for implementation.
However, the COSY ring was permanently shut down at the end of 2023 and is no longer available for experiments. 

Dedicated storage rings with only electric fields or a combination of electric and magnetic fields employing counter rotating beams are currently under investigation~\cite{abusaif2021, Anastassopoulos:2015ura, PhysRevAccelBeams.22.034001}.
In such a configuration, many of the dominant systematic effects -- arising from field imperfections, orbit distortions, and beam dynamics -- are expected to cancel out due to their opposite signs in the two oppositely circulating beams.  
This cancellation offers a powerful pathway to significantly reduce systematic uncertainties and achieve a sensitivity that could surpass current limits for hadrons, \emph{i.e.} \SI{{}e-26}{\echarge\cdot cm} for the neutron~\cite{PhysRevLett.124.081803}. 
The successful realization of such a facility would mark a significant advance in precision EDM measurements, enabling a direct test of CP violation in the hadronic sector and opening a new frontier in the search for physics beyond the Standard Model.

\textit{Acknowledgements} --- We thank the COSY crew for their dedicated support in operating the accelerator during the experiment. This work was carried out within the framework of the JEDI Collaboration and supported by an ERC Advanced Grant from the European Union (Proposal No.\,694340: \emph{Search for Electric Dipole Moments Using Storage Rings}).
The contributions of A.\,Aksentev, A.\,Melnikov, and N.\,N.\,Nikolaev were supported by the Russian Science Foundation (Grant No.\,25-72-30005).
This research also received funding from the European Union’s Horizon\,2020 research and innovation program under Grant Agreement STRONG-2020-No.\,824093.
The presented results are based on the Ph.D. thesis of A.\,Andres~\cite{Andres:999521}.

The data used in this article are openly available in Ref.~\cite{juelich_data}.


\bibliography{literature}


\section{End Matter}
\textit{Models for polarization buildup} --- 
Two models have been developed for the single-bunch and pilot-bunch methods to describe the polarization buildup and to extract the resonance strength $\epsilon$. Deviations from the ideal behavior described by Eq.~(\ref{eq:oscwf}) arise from decoherence and from deviations from the resonance frequency. 
Spin decoherence leads to a decay of the in-plane polarization, thereby reducing the effective contribution of spin rotation from horizontal to vertical polarization over time~\cite{Hempelmann:718035,PhysRevAccelBeams.27.111002}. This effect must be taken into account, in particular for finite spin-coherence times, to reliably determine the resonance strength from the initial slope or the oscillation frequency. Deviations from exact resonance account for the finite precision of the phase-lock feedback~\cite{PhysRevAccelBeams.28.062801}.

\textit{Polarization buildup using a single bunch} --- 
Decoherence is taken into account by fitting the out-of-plane angle $\alpha(t)$ of the polarization vector and the total polarization $p_\mathrm{tot}(t)$ simultaneously, where
\begin{eqnarray}
    \alpha(t) &=& \arctan(p_v(t)/p_h(t)), \\
    p_\mathrm{tot}(t) &=& \sqrt{p_v(t)^2 + p_h(t)^2},
\end{eqnarray}
where $p_h(t)$ and $p_v(t)$ denote the in-plane and the vertical polarization, respectively.
Note that the vertical polarization $p_v$ corresponds to the spin component $S_y$ in Eq.~\ref{eq:oscwf}.
Assuming an exponential decay of the in-plane polarization, characterized by
\begin{equation}
-\frac{1}{p_h(t)}\cdot\frac{\mathrm{d}p_h(t)}{\mathrm{d}t} = \mathrm{const}. = B,    
\end{equation}
this results in a non-linear term for $\alpha(t)$, which includes the constant spin kick per turn $A = 2\pi\epsilon\fc \sin\phirel$ as one of its parameters (cf. Eq. (5.13) and (5.14) in \cite{Hempelmann:718035})
\begin{equation}
\begin{aligned}
\tan&(\alpha(t)) = \\[-1mm]
&\begin{cases}
\tan(\alpha_0) e^{B t}, & A = 0, \\[2mm]
\frac{1}{2A} \left( \sqrt{E} \frac{D + \tan\left(\frac{t}{2}\sqrt{E}\right)}{1 - D \tan\left(\frac{t}{2}\sqrt{E}\right)} - B \right), & E > 0, \\[1mm]
\frac{1}{2A} \left( \sqrt{-E} \frac{D - \tanh\left(\frac{t}{2}\sqrt{-E}\right)}{1 - D \tanh\left(\frac{t}{2}\sqrt{-E}\right)} - B \right), & E < 0,
\end{cases} \\[1mm]
&\text{with } D = \frac{2A \tan(\alpha_0) + B}{\sqrt{|E|}}, \quad E = 4A^2 - B^2
\end{aligned}
\end{equation}
and
\begin{equation}
  \begin{aligned}
    &\log \left(\frac{p_{\text{tot}}(t)}{p_{\text{tot},0}}\right)= \frac{1}{2} \log \left(\frac{2 A+B \sin \left(2\alpha_{0}\right)}{2 A+B \sin(2 \alpha(t))}\right)+ \\ &\begin{cases}
    \frac{B}{\sqrt{E}} \arctan \left(\frac{\sin\left(\alpha_{0}-\alpha(t)\right) \sqrt{E}}{2 A \cos\left(\alpha_{0}-\alpha(t)\right)+B \sin\left(\alpha_{0}+\alpha(t)\right)}\right), & E>0, \\
    \frac{B}{\sqrt{-E}} \operatorname{artanh}\left(\frac{\sin \left(\alpha_{0}-\alpha(t)\right) \sqrt{-E}}{2 A \cos \left(\alpha_{0}-\alpha(t)\right)+B \sin \left(\alpha_{0}+\alpha(t)\right)}\right), & E<0.
  \end{cases}
  \end{aligned}
\end{equation}
where $\alpha_0$ and $p_{\text{tot},0}$ denote the initial out-of-plane angle and initial total polarization.

\textit{Polarization buildup using a pilot bunch} ---
This case including decoherence and off-resonance behavior is discussed in \cite{PhysRevAccelBeams.27.111002}.
The underlying formalism describes the time evolution of the polarization vector~$\vec{p}$ after a time interval $x$, described in terms of a dimensionless variable,
\begin{equation}
x = 2\pi \epsilon_{\text{m}} (n - n_{\text{WF}}^{\text{on}})
\; \text{or} \;
x = 2\pi \epsilon_{\text{m}} \fc (t - t_{\text{WF}}^{\text{on}}),
\end{equation}
where $n$ denotes the turn number and $\epsilon_{\text{m}}$ the measured resonance strength. Here, $x_{\text{WF}}^{\text{on}} = 2\pi \epsilon_{\text{m}} n_{\text{WF}}^{\text{on}}$ corresponds to the time when the rf Wien filter is switched on.
The polarization vector is expressed as
\begin{equation}
  \vec{p}(x) = \mathbf{E}(x) \cdot \vec{p}(x = x_{\text{WF}}^{\text{on}}),
\end{equation}
with $\vec{p} = (p_{\text{r}}, p_{\text{v}}, p_{\text{t}})^{\text{T}}$.
The vertical component $p_{\text{v}}$ is oriented along the invariant spin axis~$\vec{n}$, the tangential component $p_{\text{t}}$ is defined along the beam direction, and the radial component $p_{\text{r}}$ lies perpendicular to both $\vec{n}$ and $p_{\text{t}}$.
The polarization transfer matrix $\mathbf{E}(x)$ is given by Eq.~(91) in \cite{PhysRevAccelBeams.27.111002}. 
The initial polarization condition for the experiment is defined by
\begin{equation}
  \vec{p}(x = x_{\text{WF}}^\text{on}) =  
  \left(\begin{array}{c}
    p_{\text{h},0}\sin(\phirelzero)\\
    p_{\text{v},0}\\
    p_{\text{h},0}\cos(\phirelzero)
  \end{array}\right),
\end{equation}
where $\phirelzero$ denotes the relative phase at the time the rf Wien filter is switched on. The initial in-plane polarization is given by $p_{\text{h},0}^2=p_{\text{r},0}^2+p_{\text{t},0}^2$, and $p_{\text{v},0}$ denotes the initial vertical polarization, which can be non-zero if the rotation of the vertical polarization into the horizontal plane by the rf solenoid is incomplete. 
While the vertical polarization remains preserved over time, an exponential decay of the in-plane polarization is assumed. The decoherence parameter $Q$, related to the Spin Coherence Time $\tau_{\text{SCT}}$, is defined as 
\begin{equation}
Q=\frac{1}{2\pi\epsilon_{\text{m}}\fc \tau_{\text{SCT}}}.
\end{equation}

The parameter $\delta$ characterizes the off-resonance behavior and is defined as
\begin{equation}  
\delta = \frac{2\pi(f_s - f_{\text{WF}})}{\fc}.  \label{eq:off_resonance_delta}
\end{equation}
In contrast to these corrections for the Spin Coherence Time, a detuned rf-Wien-filter frequency leads to a systematic shift of the measured resonance strength. 
The measured and the detuned resonance strength are connected via the dimensionless parameter $\rho$
\begin{equation}
\sin(\rho) = \frac{\epsilon}{\epsilon_{\text{m}}} \text{ and } \cos(\rho) = \frac{2\delta}{4\pi\epsilon_{\text{m}}}, \label{eq:off_resonance_behaviour}
\end{equation}
which allows for the determination of the unbiased resonance strength.
The tangential, radial and vertical components of the polarization are given by
\begin{equation}
  \begin{aligned}
    p_\text{t} = &-e^{-Qx}\cos(\rho)\sin(x)\cos(\phirelzero)p_{\text{h},0}\\
          &+e^{-Qx}\sin(\rho)\sin(x)p_{\text{v},0}\\
          &+e^{-Qx}\cos(x)\sin(\phirelzero)p_{\text{h},0},
  \end{aligned}
\end{equation}
\begin{equation}
  \begin{aligned}
    p_\text{r} = & \left[e^{-2Qx}\sin(\rho)^2+e^{-Qx}\cos(\rho)^2\cos(x)\right]\cos(\phirelzero)p_{\text{h},0}\\ 
          &+ \left(e^{-2Qx}-e^{-Qx}\cos(x)\right)\cos(\rho)\sin(\rho)p_{\text{v},0} \\
          & + e^{-Qx}\cos(\rho)\sin(x)\sin(\phirelzero)p_{\text{h},0},
  \end{aligned}
\end{equation}
and
\begin{equation}
  \begin{aligned}
    \pv = & -e^{-2Qx}\cos(\rho)\sin(\rho)\sin(\phirelzero)p_{\text{h},0}\\ 
            & +e^{-Qx}\cos(\rho)\sin(\rho)\cos(x)\sin(\phirelzero)p_{\text{h},0}\\ 
            & + \left[e^{-2Qx}\cos(\rho)^2+e^{-Qx}\sin(\rho)^2\cos(x)\right]p_{\text{v},0} \\
            & - e^{-Qx}\sin(\rho)\sin(x)\sin(\phirelzero)p_{\text{h},0}.
  \end{aligned}
  \label{eq:pilot_bunch_vertical_component}
\end{equation}

\begin{figure}[t!]
    \begin{center}
     \includegraphics[width=\linewidth]{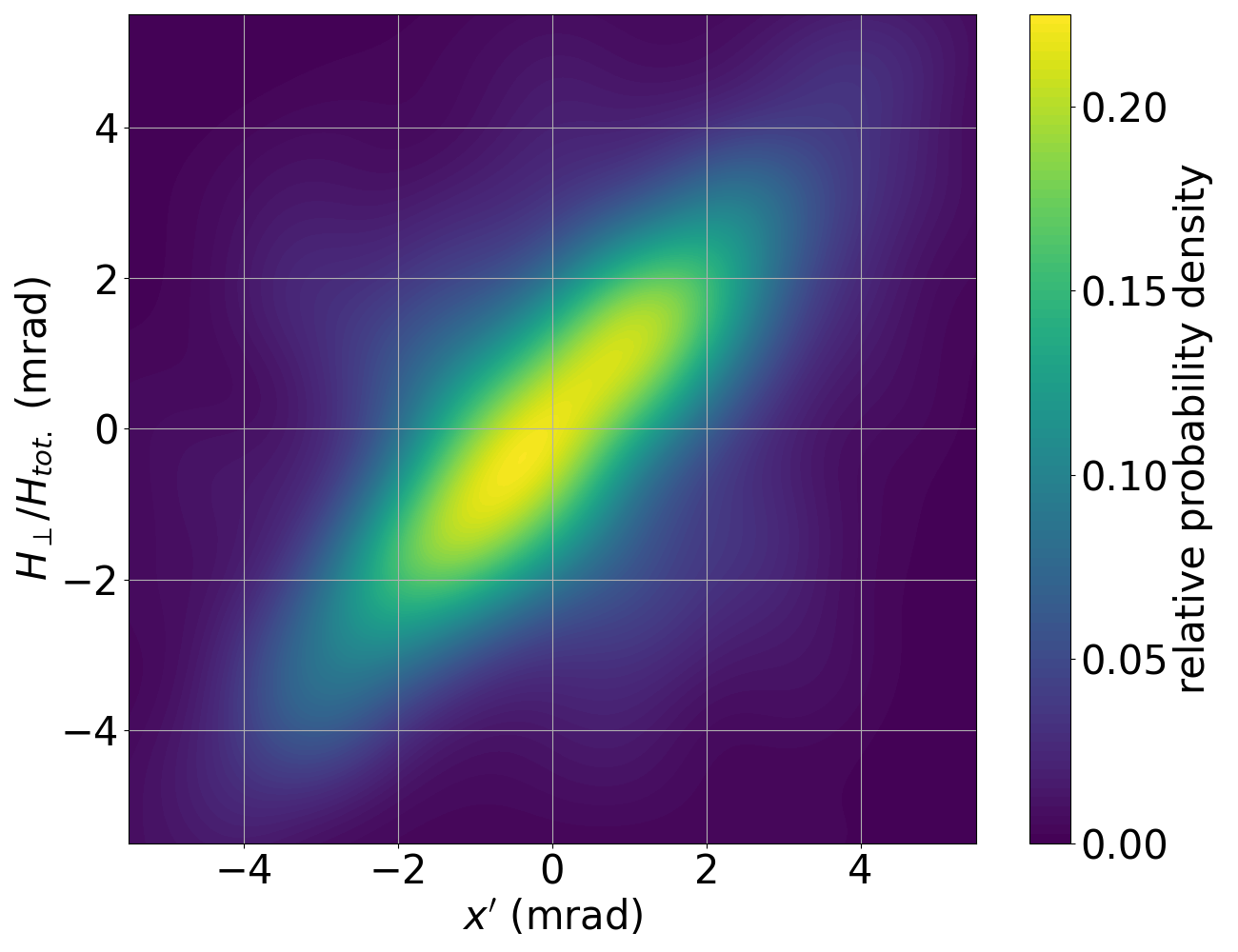}
        \caption{Perpendicular contributions to the main (vertical) magnetic field as a function of a rotation of the beam axis in the horizontal plane ($x^\prime$), illustrating the sensitivity of the field orientation to uncertainties in the beam parameters.
        The plot resembles a projection of the 16-parameter space onto the $x^\prime$ axis and is based on 800 full-wave simulations, further processed using a machine-learning-based sparse polynomial chaos expansion as discussed in Ref.~\cite{Slim:2016dct}. Expected uncertainties in $x^\prime$ are in the order of \SIrange[range-phrase={--}, range-units=single]{1}{2}{mrad}.
        The colors indicate the probability for the $H_{\perp}$ contribution given at a certain $x'$.
        \label{fig:rf_wf_sys}}
    \end{center}
\end{figure}

Using the polarimeter, only the in-plane polarization 
\begin{equation}
  \ph = \sqrt{p_\text{r}^2+p_\text{t}^2}, \label{eq:pilot_bunch_in_plane_polarization}
\end{equation}
and the vertical polarization can be measured. From the event rates in the polarimeter quadrants, the left-right ($\epsilon_{LR}$) and up-down asymmetries ($\epsilon_{UD}$) are determined, which scale directly with the vertical and horizontal polarization, respectively:
\begin{align}
    \ph &\rightarrow \epsilon_{\text{UD}}, \\
    \pv &\rightarrow \epsilon_{\text{LR}}.
\end{align}
Finally, both $p_\text{v}$ and $p_\text{h}$ are fitted simultaneously to the measured asymmetries and the resonance strength $\epsilon$ can be extracted for one map point.

\textit{Systematic Uncertainties} ---
Beam- and spin-tracking simulations accounting for misalignments of the ring magnets and beam-position monitors have been carried out. In addition, simulations employing uncertainties in the envisaged final rf Wien filter configuration were performed to assess their impact on the field orientation. 
Originally, an additional experimental run using this final rf Wien filter geometry, including the planned ferrite components, had been foreseen. This could not be realized, as COSY was shut down at the end of 2023, excluding also further dedicated measurements to experimentally constrain sources of systematic uncertainties. 

Moreover, the Rogowski coils developed as beam position monitors at the entrance and exit of the rf Wien filter~\cite{rogowski_bpm} had been commissioned during the beam time used for the measurements presented in this work. Consequently, local beam orbit monitoring within the rf Wien filter was limited, and beam position information relied on BPMs located further upstream and downstream. 
As a result, possible deviations from the optimal beam orbit within the rf Wien filter could not be excluded.
To address these limitations, additional simulations on the rf Wien filter fields were performed following the approach of Ref.~\cite{Slim:2016dct}.
In total 16 parameters were varied. 
These simulations included variations of both the rf Wien filter geometry and the beam parameters, including horizontal and vertical translations and rotations of the beam. The dominant contribution was found to originate from a rotation of the beam axis in the horizontal plane, as illustrated in Fig.~\ref{fig:rf_wf_sys}. 
This could also explain the observed map-to-map variations, if one allows for map-dependent orbit changes in the rf Wien filter.

Combining all identified sources of systematic uncertainty, the resulting uncertainty in the magnetic-field orientation and thus in the direction of the invariant spin axis can reach the level of a few milliradians, consistent with the observed systematic offset of the final result.

\end{document}